\documentclass{article}
\begin{document}
\begin{center}
{{\bf \Large Research News --- Progess in determination of
neutrino oscillation parameters}}\footnote{
In Memoriam: John N. Bahcall (1934-2005), Raymond Davis Jr.
(1914-2006)}

\bigskip

B. Ananthanarayan$^a$\\ 
Chitra Gautham$^b$\\
Aquila Mavalankar$^c$\\
K. Shivaraj$^a$\\
S. Uma Sankar$^d$\\
A. Upadhyay$^a$

\bigskip
{\small
$^a$ Centre for High Energy Physics, Indian Institute of Science,
Bangalore 560 012\\
$^b$ Madras Christian College, Tambaram, Chennai 600 059\\
$^c$ St. Stephen's College, New Delhi 110 007\\
$^d$ Department of Physics, Indian Institute of Technology, Bombay 400 076
}
\medskip
\end{center}

\begin{abstract}
Recent results from the MINOS experiment at Fermilab reconfirm
neutrino oscillations.  We describe briefly this experiment and
discuss how this and other experiments enable us to determine
fundamental parameters of elementary particle physics in the
neutrino sector.
\end{abstract}


A press release dated March 30, 2006 from the US laboratory Fermilab 
reported the first results from a new neutrino experiment, MINOS
(Main Injector Neutrino Oscillation Search) \cite{minospr}. The Main 
Injector accelerator at Fermilab produces an intense beam of muon 
neutrinos and directs them at the MINOS detector in the Soudan mine,
at a depth of 716 meters, 
in Minnesota, 732 km away. The MINOS experiment expected to observe
$177 \pm 11$ events but observed only $92$, which is $7.5 \sigma$
away from $177$. The probability that random fluctations are responsible
for this shortfall is only part in $10^{10}$. Thus MINOS experiment
confirmed that there is a shortfall in the number of muon neutrinos
if they are detected a long distance away from their point of production. 

Two water Cerenkov detectors, Irvine-Michigan-Brookhaven (IMB)
\cite{imb} and Kamiokande \cite{kamio} have first observed this 
deficit fifteen years ago in muon neutrinos produced in the atmosphere.
The interaction rate for downward going neutrinos, which travel distances 
of the order of 100 km, was consistent with expectation but the rate for
upward going neutrinos, which travel thousands of km, was only about
$60\%$ of the expectation. The field of neutrino physics received a 
tremendous boost in 1998, when the results of the very large water Cerenkov 
detector, Super-Kamiokande were announced \cite{sk}. These results not only 
confirmed the deficit of atmospheric muon neutrinos when they travel long 
distances, but also showed that there are definite correlations between the 
amount of deficit and the distance travelled and also between the 
amount of deficit and the neutrino energy. Both the deficit and the
correlations can be explained by assuming that muon neutrinos, during 
their long travel, {\it oscillate} into another type of neutrino.

However, there are still large uncertainties in the calculation of 
atmospheric neutrino fluxes. The long baseline neutrino experiments
were planned to find evidence for the spectral distortion caused by
oscillations by locating the minimum of the muon neutrino survival 
probability. The location of this minimum can be used to determine
the neutrino oscillation parameters accurately. 
In Super-Kamiokande, the observed neutrinos have varying energies and
also varying pathlengths. Locating the minimum of the survival
probability is very difficult to do in Super-Kamiokande because the 
data sample spans large ranges in energis and pathlengths.
The first long baseline neutrino experiment, called K2K (KEK to 
Super-Kamiokande 230 km away) in Japan, was a low statistics experiment 
which confirmed the muon neutrino deficit. It also observed a distortion 
in the energy spectrum of the muon events, that is consistent with   
oscillation hypothesis. The number of events 
in recent MINOS results also is not very large and the current 
accuracy of MINOS is only slightly better than that of K2K.
But it is expected that with about 5 years of data taking, they will
increase their number of events twenty fold. This will enable them to  
observe the minimum of the survival probability and determine the 
neutrino oscillation parameters precisely.  

If the MINOS results are interpreted in terms of three flavour
neutrino oscillations, they give the following values for neutrino
oscillation parameters:
$\Delta m^2_{23}=(3.05^{+0.60}_{-0.55}\pm 0.12)\cdot 10^{-3}
{\rm eV}^2$,
and $\sin^2(2\theta_{23})=0.88^{+0.12}_{-0.15}\pm 0.06$.
It can be seen from the results presented in \cite{minospr,schwetz}
that the MINOS results have already improved the precision in
the determination of $\Delta m^2_{23}$ but the precision in $\theta_{23}$
is still controlled by atmospheric neutrino data.

Here we describe the MINOS experiment in a little more detail. 120 GeV 
protons are extracted from Fermilab's Tevatron Main injector at the rate 
of $10^{13}$ protons per second and these protons are directed to a 
fixed graphite target. Pions and Kaons are selected from the 
resultant spray of secondary particles and are focused into a 
675 m evacuated decay pipe where they decay into muons and muon neutrinos. 
Muons are absorbed by 200 m rock, leaving behind a pure beam of 
muon neutrinos. The near detector (aprox 1 kton) is located 300 m 
down from the hardron absorber. The far detector is placed in 
Soudan and is a 5400 ton detector. It was started on March 4, 2005 
and it aims to study neutrinos in the energy range
1-30 GeV and to provide a 
more precise measurement of the mass difference and mixing angle
responsible for the disappearance of atmospheric neutrinos 
($\Delta m^2_{23}$ and $\theta_{23}$ mentioned above). 
The experiment also studies $\nu_\mu \to \nu_e$ 
oscillations in the same exposure. 
The magnetized steel 
plates used in detectors allow us to distinguish $\nu_\mu$
CC events from $\bar{\nu}_\mu$
CC events, and thus to search for CPT violation in 
atmospheric neutrino oscillations. 

We also briefly recall the results from K2K.
K2K which is a long-baseline experiment where the 
neutrinos are produced in KEK using a proton synchroton which accelerates 
protons to an energy of 12 GeV. These protons strike an aluminum target 
and produces pions which then decay into muons and neutrinos. A detector in 
KEK first detects the neutrino flux before the neutrinos travel 250 kms to 
SuperKamiokande, where they are detected by the Cerenkov principle in 
50,000 tons of water. In the absence of neutrino oscillations, 
there were supposed to be $158.1^{+9.2}_{-8.6}$
events, but only 112 were observed \cite{k2k}.  

Neutrinos are elementary particles, which are neutral counterparts
of the charged leptons, namely the electron, the muon and the
$\tau$-lepton all of which participate in the weak interactions.
The determination of neutrino
properties remains notoriously difficult from the point of view of
experiments and their precise determination remains one of the great
challenges and goals of elementary particle physics research today.
At the moment, there is no information of even the values of their
individual masses, which are today bounded by experiments as follows 
\cite{pdg}: $m_1< 3 {\rm eV},\, m_2<190 {\rm keV}, m_3<18.2 {\rm MeV}$. 
It is worth noting here that the 
direct detection of the $\nu_\tau$ was 
reported for the first time only as recently as 
2000 \cite{DONUT} by the
Fermilab DONUT
(Direct Observation of Nu Tau, E872) experiment.
This experiment used protons accelerated by the Tevatron to produce a
$\nu_\tau$ beam and an active emulsion detector.

We note here that the subject of neutrino physics is considered 
so important at the present time 
that the American Physical Society
has commissioned a multi-divisional study whose
reports are now available to the public on the internet~\cite{aps}.
In particular, several working groups were formed, on
solar and atmospheric experiments, reactor experiments, neutrino
factory and beta beam experiments, neutrinoless double beta decay
and direct search experiments, neutrino astrophysics and cosmology
and on theory.  

The presence of neutrino
oscillations implies existence of distinct and non-vanishing masses
for the two heavier flavours. In particular, there are now three
masses $m_1$, $m_2$ and $m_3$ and three angles that 'mix' the
neutrino flavours denoted by $\theta_{12}$, $\theta_{23}$ and
$\theta_{13}$(This last parameter has not been experimentally
determined so far. A reactor neutrino experiment, CHOOZ, which measured 
the survival probability of electron anti-neutrinos from a nuclear power 
reactor in France, observed no deficit and set the bound 
$\sin^2\theta_{13}<0.05$ \cite{CHOOZ}).  
In addition, neutrinos may also be Majorana particles, that is
they are neutral fermions which are their own antiparticles.
In that case there would be also the possibility of lepton number
violation in nature by two units, $\Delta L=2$, which, e.g.,
is the basis of the neutrinoless double-beta decay
experiment. The matrix that describes the
neutrino mixing called the PMNS matrix, named for Pontecorvo, Maki,
Nakagawa and Sakata after the authors who first described it.
In form, it is very similar to its counterpart in the quark sector, 
namely the CKM matrix named for Cabibbo, Kobayashi and Maskawa
\cite{pdgckm}. However, the angles in the CKM matrix are all well
measured and all three of them are small (less than $12^\circ$) 
whereas two of the three mixing angles in neutrino sector are 
larger than $30^\circ$, showing that the patterns of mixing in
quark and lepton sectors are very different. Note that the CKM matrix
has a non-zero phase $\delta$ which leads to CP violation in the decays
of K and B mesons and is well determined from the experimental 
measurement of CP violating observables. CP violation may also occur
in neutrino oscillations if PMNS matrix contains a non-zero phase.
Such a non-zero phase is possible in the mixing of three neutrinos
(as it is in the case of three quarks). Such a phase leads
to the prediction that the oscillation probabilities for neutrinos 
and anti-neutrinos are unequal. One can establish CP violation in
neutrino sector by studying $\nu_\alpha \rightarrow \nu_\beta$
and $\bar{\nu}_\alpha \rightarrow \bar{\nu}_\beta$, where $\alpha$
and $\beta$ are neutrino flavours. CPT violation in neutrino sector
can be established if one can show that the survival probabilities
$P(\nu_\alpha \rightarrow \nu_\alpha)$
and $P(\bar{\nu}_\alpha \rightarrow \bar{\nu}_\alpha)$ are unequal.

We now turn to the solar neutrino problem which has 
been discussed earlier~\cite{prev}. This problem also
is resolved by postulating neutrino oscillations. 
To recall briefly the salient features of this problem, 
the measured fluxes of solar neutrinos at the radio chemical experiments
at Homestake (associated with Raymond Davis Jr.), 
SAGE, Gallex and GNO experiments were significantly
lower than those predicted by the standard solar model (SSM, associated
with John N. Bahcall). 
The experiment at Homestake was based on capture of neutrinos
by chlorine nuclei, while those at the other three were based
on capture of neutrinos by gallium nuclei.  That there was
a solar neutrino problem was also confirmed by 
Kamiokande and SuperKamiokande experiments, which detected 
solar neutrinos in {\it real time} by observing neutrino-electron
scattering. According to standard electroweak model (SEM), the sun
should emit only electron neutrinos which can be detected on earth
through their charged current (CC) interactions. In such an interaction,
the electron neutrino exchanges charge with another particle and produces 
an electron. The observed deficit of solar neutrinos, compared to the
predictions of SSM is explained by assuming that electron neutrinos
are oscillating into muon/tau neutrinos. These neutrinos cannot be 
observed in the above detectors because their CC interactions produce
a muon/tau in their final states. The low energies of solar neutrinos
preclude the production of these heavy mass particles. Thus observed
fluxes were lower than the predicted fluxes. 

This raises the question: Is it at all possible to observe the muon/tau
neutrino fraction in the solar neutrinos reaching the earth? It is 
possible to observe them through their neutral current (NC) interactions.
In a NC interaction, a neutrino of a given flavour retains its identity
and exchanges only energy with the other participating particle. If
heavy water is used as a target, then a neutrino interacting with a
deuteron will break it up and the resultant neutron can be detected.
Sudbury Neutrino Observatory (SNO) used this technique to measure
the NC interaction rates of solar neutrinos. According to SEM, 
the NC interaction rates of all three types of neutrinos, with any given 
particle, are the same. Thus, if the electron neutrinos emitted by
the sun are changing into muon/tau neutrinos then the NC interaction
rate should be consistent with the predictions of SSM. The first results
of SNO, announced in 2001 \cite{sno}, definitely established that the NC 
interaction rates of solar neutrinos are indeed in accordance with 
SSM predictions. SNO independenly measured the CC interaction rate 
of electron neutrinos also, by detecting the final state electrons. 
These measurements are consistent with the previous measurements and
show that there is indeed a shortfall of electron neutrinos from the
sun as they arrive on earth. Detection of the neutron in SNO is subject
to various uncertainties. Therefore SNO decided to use three different
techniques to detect the neutron. In the first phase, they used the 
heavy water itself to detect the neutron. In the recently concluded
second phase, salt was added to the heavy water to improve the neutron
detection efficiency~\cite{saltphase}.  
In the current third phase, the salt is withdrawn 
and $^3He$ proportional counters added which detect neutrons through
the reaction $n+^3He\to p+^3H$. 
This will assist in reducing the error in the measurement of the 
NC reactions rates.

The combined results of the all solar neutrino experiments require that
the resolution of the solar neutrino problem is through neutrino
oscillations with a large mixing angle (LMA) modified by the   
MSW (Mikheyev-Smirnov-Wolfenstein) mechanism.
Recalling briefly here, the MSW
effect results from the enhancement of the flavour oscillation
of $\nu_e$ generated in various nuclear reactions in
the solar interior to
$\nu_\mu/\nu_\tau$, due to their interaction
with the dense solar interior.  The mechanism turns out to be
efficient for the energy range of the detected neutrinos and
is known as 'resonance conversion'. 
The final determination of solar neutrino oscillation parameters are:
$\Delta m_{12}^2=8^{+0.6}_{-0.4}\cdot 10^{-5}{\rm eV}^2$,
$\theta_{12}=(33.9^{+2.4}_{-2.2})^\circ$.
It may be noted that results coming from the BOREXINO
experiment, which seeks to measure the intensity of 
the monochromatic 861 keV neutrino line could play an important
role in constraining solar models and neutrino oscillation
models. An ambitious project based on the 1976 proposal of
R. S. Raghavan that is being planned is the LENS (Low Energy
Neutrino Spectroscopy) experiment, which is based on a radiochemical
reaction involving Indium targets and can also measure low-energy
neutrinos produced in the solar proton-proton cycle in real time. 

In order to establish the validity of the conclusions from the SNO
observations, {\it viz.}, the resolution of the solar neutrino
problem through the LMA-MSW solution, the KamLAND experiment 
(Kamioka Liquid scintillator Anti-Neutrino Detector) was set up in
Toyama, Japan in 2002. The detector 
is surrounded by 53 Japanese commercial power reactors (~180 kms away), 
which produce $\bar{\nu}_e$. The neutrinos are detected using the Cerenkov 
principle in 1000 tons of mineral oil, benzene and fluorescent chemicals. 
Without neutrino oscillation the experiment expected to see 
$86.6\pm 5.6$ events. However only 54 events were observed. 
This confirmed the picture of LMA oscillations in vacuum
\cite{kamland}.

Finally, the presence of non-vanishing masses
for the neutrinos is likely to hold the key to our understanding of
not just the properties of elementary particles, but also to the
entire history of the Universe.  As a result, it is important to
demonstrate neutrino oscillations in several different settings,
and to independently measure mass square differences. That is
why the results of MINOS and the future experiments T2K (under
construction) and NOVA and INO (in planning stages) are important.
Note that the above experiments, which detect neutrino oscillations,
can only give information on neutrino mass differences but not on
the mass of the lightest neutrino. This information is expected to
come from tritium beta decay experiments \cite{tritium} and 
double-beta decay experiments \cite{doublebeta}.

We recall here the results from a somewhat
controversial experiment known as LSND 
(Liquid Scintillator Neutrino Detector)~\cite{LSND}.
The experiment observed excesses of events for both the 
$\bar{\nu}_\mu\to \bar{\nu}_e$ and $\nu_\mu \to \nu_e$ oscillation 
searches. If confirmed, the results of LSND experiment imply
that there is a mass-squared difference of about $0.1$ eV$^2$
between two neutrino mass eigenstates. This conclusion, inevitably
leads to the prediction that there should be a fourth light 
neutrino, because we can't have three mass-squared differences 
of three different orders of magnitude with just three neutrinos.
The fourth neutrino must be sterile (essentially non-interacting)
because the results of the
LEP experiments have shown that there are only three light neutrinos
with the standard interactions \cite{lep}. MiniBooNE 
(Mini Booster Neutrino Experiment) \cite{miniboone} experiment at Fermilab 
has been taking data and it can confirm or rule out the LSND result. 
It consists of a 1 GeV neutrino beam from pion decay and a single, 
800-ton mineral oil (12-meter diameter sphere) detector. 
The MiniBooNE detector is located 500 meters downstream of the neutrino source.
The MiniBooNE results are expected to be announced in a year's time. 

In the following section we discuss experiments whose goal is 
neutrino astronomy, identifying astrophysical sources of neutrinos 
which include medium energy (about 100MeV) sources within our galaxy 
like supernovae and the sun and high energy (a few GeV and greater) 
extragalactic sources like exploding stars, gamma ray bursts, black holes 
and neutron stars. Some of these experiments 
also try to detect neutrino oscillations.

It may be recalled that the field of supernovae neutrino physics
was born with the detection of the neutrinos from the supernova 1987A
by IMB \cite{imbsn} and Kamiokande-II \cite{kamiosn}.
Some of the current neutrino experiments, Superkamiokande and SNO,
are capable of observing supernova neutrinos. There are various 
plans to improve the current experiments or design new experiments
to enhance the detection capabitilies of supernova neutrinos. 
Study of signals from supernova neutrinos in various different
types of detectors is an important topic in neutrino physics today
\cite{amol,imsc}.

Neutrino telescopes can be located under water. 
DUMAND (Deep Underwater Muon And Neutrino Detector) and Baikal were 
among the first to explore the unchartered waters. The former was 
located offshore Hawaii in the Pacific Ocean, while the latter was 
located in Lake Baikal in Siberia. The next one is NESTOR 
(Neutrino Extended Submarine Telescope With Oceonographic Research) 
located offshore Greece, in the Mediterranean Sea, which 
detects neutrinos in the TeV range. It determined the cosmic ray 
muon flux as a function of the zenith angle and the energy spectrum and 
composition of primary cosmic rays. The future experiments include 
ANTARES (Astronomy with a Neutrino Telescope and Abyss environmental RESearch) 
located offshore France and NEMO (NEutrino Mediterranean Observatory) 
offshore Sicily. ANTARES is a step towards a cubic kilometre 
telescope in the Mediterranean Sea. ANTARES and NEMO are designed to 
detect high energy neutrinos in the TeV to PeV range.

The new generation of neutrino experiments also includes the 
Antarctic Muon and Neutrino Detector Array (AMANDA) built to detect 
extra-solar sources of neutrinos. 
The photomultiplier 
tube (PMT) array located between 1500 m and 2000 m of Antarctic ice detects 
neutrinos coming through the earth from the northern sky using the 
Cerenkov principle. The depth of the array effectively blocks 
atmospheric neutrinos. 
Compared to underground detectors like 
SuperKamiokande,
AMANDA is capable of looking at higher energy neutrinos ($>50$ GeV). AMANDA 
demonstrated the viability of a neutrino telescope in ice 
and in 2005 was officially incorporated into its successor 
ICECUBE. It will encompass a cubic kilometre of ice and will 
be completed in 2010-11. IceCube will be able to 
explore the PeV ($10^{15}$ eV) energy region. The detection of 
cosmic neutrino beams would open the opportunity to study 
neutrino oscillations over Megaparsec baselines.

We take this opportunity to mention that there is a proposal to
build a neutrino observatory in India. It is called India-based
Neutrino Observatory (INO) and is likely to be built in Nilgiris.
It is a 50 kiloton detector consisting of magnetized iron sheets
interspersed with active detector elements. It can detect muons
produced by atmospheric muon neutrinos and can determine its 
charge, energy and direction accurately. This will enable the 
detector to measure the atmospheric mass-squared difference 
very accurately (to about $10 \%$) \cite{ino}. 
In addition, it can also determine the hierarchy of neutrino mass
eigenstates \cite{gandhi} and the deviation from maximality of
$\theta_{23}$ \cite{ChoubeyRoy} if the CHOOZ mixing angle $\theta_{13}$
is large enough.

We conclude by remarking here on the constraints that the present day
neutrino experiments and the future experiments will place on theoretical
models for neutrino masses and mixings.  One of the spectacular
possibilities is that of 'grand unification' of the standard model
interactions, {\it viz.} the strong and electro-weak interactions.
Such models are often based on larger 'gauge groups' into which the
gauge symmetries of the standard model would fit into.  Notable among
these is $SO(10)$ unification, which would naturally accomodate
a right handed neutrino, and would also admit both Dirac and Majorana
masses for neutrinos. In this event, the well-known 'see-saw' mechanism
could generate the observed spectrum of masses \cite{mink,seesaw}. 
There are many
models today which predict and accomodate the observed masses and
mixings. The forthcoming experimental results will help us in discriminating 
among these models.

\bigskip

\noindent
{\bf Update:}  After this article was written, 
in a press release dated August 7, 2006 from Fermilab~\cite{Minosupdate}, 
it has been     
reported that the MINOS collaboration has analyzed data representing 37%
increase compared to the results presented end of March 2006.  In the           
absence of neutrino oscillations, they would have expected to see 
$336\pm 14$ muon neutrinos, but instead they record 215 muon neutrino 
events, thereby       
confirming their previous results,
with the best fit for $\Delta m_{23}^2=(2.74^{+0.44}_{-0.26})\cdot
10^{-3} {\rm eV}^2$.

\bigskip

\noindent{\bf Acknowledgement:} We thank Srubabati Goswami for
a careful reading of the manuscript and useful comments.
Thanks are also due to Rabi Mohapatra and Probir Roy for comments
and references.


\newpage

\begin{thebibliography}{99}

\bibitem{minospr}
See the website
http://www.fnal.gov/pub/presspass/press\_releases/minos\_3-30-06.html
\bibitem{imb}
D. Casper {\it et al}, [IMB Collaboration],
Phys. Rev. Lett. {\bf 66}, 2561 (1991).
\bibitem{kamio}
K. Hirata {\it et al}, [Kamiokande Collaboration]
Phys. Lett. {\bf B280}, 146 (1992).
\bibitem{sk}
Y. Fukuda {\it et al}, [Super-Kamiokande Collaboration],
Phys. Rev. Lett. {\bf 81}, 1562 (1998), {\it ibid} {\bf 82}, 2644 (1999).
\bibitem{schwetz}
T. Schwetz, hep-ph/0606060
\bibitem{k2k}
M. Ahn {\it et al}, [K2K Collaboration], hep-ex/0606032
\bibitem{pdg}
S. Eidelman {\it et al}, Review of Particle Properties, 
Phys. Lett. {\bf B492}, 442 (2004).
\bibitem{DONUT}
  K.~Kodama {\it et al}, [DONUT Collaboration],
  Phys. Lett. {\bf B504}, 218 (2001).
\bibitem{aps}
  http://www.aps.org/neutrino
\bibitem{CHOOZ}
M. Apollonio {\it et al}, [CHOOZ Collaboration],
Phys. Lett. {\bf B466}, 415 (1999).
\bibitem{pdgckm}
S. Eidelman {\it et al}, Review of Particle Properties, 
Phys. Lett. {\bf B492}, 130 (2004).
\bibitem{prev}
  Ananthanarayan, B. and Singh, R.K., Curr.\ Sci.\ {\bf 83}, 553 (2002)\\       
  Indumathi, D., Curr.\ Sci.\ {\bf 85}, 1662 (2003)
\bibitem{sno}
  Q.~R.~Ahmad {\it et al.}  [SNO Collaboration],
  Phys.\ Rev.\ Lett.\  {\bf 87}, 071301 (2001)
  [arXiv:nucl-ex/0106015];
  Phys.\ Rev.\ Lett.\  {\bf 89}, 011301 (2002)
  [arXiv:nucl-ex/0204008];
  Phys.\ Rev.\ Lett.\  {\bf 89}, 011302 (2002)
  [arXiv:nucl-ex/0204009].
\bibitem{saltphase}
SNO Collaboration: B. Aharmim {\it et al}, Phys. Rev. C{\bf 72},
055502, 2005;
See also
A. Bandyopadhyay, S. Choubey, S. Goswami, S. T. Petcov and D. P. Roy,
Phys. Lett. {\bf B583}, 134 (2004).
\bibitem{kamland}
KamLAND Collaboration: K. Eguchi {\it et al}, Phys. Rev. Lett.
{\bf 90}, 021802, 2003; 
KamLAND Collaboration: T. Araki {\it et al}, Phys. Rev. Lett.
{\bf 94}, 081801, 2005; 
\bibitem{tritium}
C. Kraus {\it et al}, Eur. Phy. J. {\bf C 40}, 447 (2005)
\bibitem{doublebeta}
S. M. Bilenky, hep-ph/0509098.
\bibitem{LSND}
A. Aguilar {\it et al}, Phys. Rev. D{\bf 64}, 112007 (2001).
\bibitem{lep}
LEP Collaborations, hep-ex/0412015, Phys. Lett. {\bf B 276}, 247 (1992).
\bibitem{miniboone}
H. L. Ray for {MiniBoone Collaboration}, hep-ex/0411022
\bibitem{imbsn}
  C.~B.~Bratton {\it et al.}  [IMB Collaboration],
  Phys.\ Rev.\ D {\bf 37}, 3361 (1988).
\bibitem{kamiosn}
  K.~Hirata {\it et al.}  [KAMIOKANDE-II Collaboration],
  Phys.\ Rev.\ Lett.\  {\bf 58}, 1490 (1987).
\bibitem{amol}
A. S. Dighe and A. Yu. Smirnov, Phys. Rev. D{\bf 62}, 033007 (2000)
\bibitem{imsc}
G. Datta, D. Indumathi, M. V. N. Murthy and G. Rajasekaran,
Phys. Rev. D{\bf 62}, 093014 (2000), 
Phys. Rev. D{\bf 64}, 073011 (2001). 
\bibitem{ino}
See: http://www.imsc.res.in/\~{~ino}
\bibitem{gandhi}
R. Gandhi {\it et al}, Phys. Rev. Lett. {\bf 94}, 051801 (2005);
Phys. Rev. D{\bf 78}, 053001 (2006).
\bibitem{ChoubeyRoy}
  S.~Choubey and P.~Roy,
  Phys.\ Rev.\ D {\bf 73}, 013006 (2006).
\bibitem{mink}
P. Minkowski, Phys. Lett. {\bf B 67}, 421 (1977).

\bibitem{seesaw}
 M.~Gell-Mann, P.~Ramond, and R.~Slansky, \emph{Supergravity}
(P.~van Nieuwenhuizen et al. eds.), North Holland, Amsterdam,
1980, p.~315; T.~Yanagida, in \emph{Proceedings of the
Workshop on the Unified Theory and the Baryon Number in the
Universe} (O.~Sawada and A.~Sugamoto, eds.), KEK, Tsukuba, Japan,
1979, p.~95; S.~L. Glashow, \emph{The future of elementary
particle physics}, in \emph{Proceedings of the 1979 Carg{\`e}se Summer
Institute on Quarks and Leptons} (M.~L{\'e}vy et al. eds.), Plenum
Press, New York, 1980, pp.~687--71;R.~N.~Mohapatra and G.~Senjanovic,
  Phys.\ Rev.\ Lett.\  {\bf 44}, 912 (1980).




\bibitem{Minosupdate}
http://www.fnal.gov/pub/today/archive\_2006/today06-08-07.html
\end{thebibliography}
\end{document}